%% file: EDTM_arxiv.tex
\documentclass[12pt,a4paper]{article} 
\usepackage[left=1.6cm, right=1.6cm, bottom=2.54cm, top=1.9cm]{geometry}
\usepackage{amsmath,amssymb,amsfonts}
\usepackage{algorithmic}
\usepackage{textcomp}
\usepackage{xcolor}
\usepackage{caption}
\usepackage{amsmath}
\usepackage{tikz, graphicx}
\usepackage{epsfig}
\usepackage{amssymb}
\usepackage{multirow}
\usepackage{float}
\usepackage{bm}
\usepackage{color}

\usepackage[lofdepth,lotdepth]{subfig}
\usepackage{ulem}
\usepackage{xfrac}
\usepackage{bigstrut}
\usepackage{multicol}
\usepackage{bigints}
\usepackage[noadjust]{cite}
\usepackage{tabularx}
\usepackage{booktabs}
\usepackage{array}
\usepackage{lettrine}
\newcolumntype{+}{>{\global\let\currentrowstyle\relax}}
\newcolumntype{^}{>{\currentrowstyle}}

\include{macros}

\begin{document}
\pagenumbering{gobble}

\title{Ferroelectric based FETs and synaptic devices for highly energy efficient computational technologies}
\author{ \large{ D. Esseni, R. Fontanini, D. Lizzit, M. Massarotto, F. Driussi, M. Loghi }\\
\small{DPIA, University of Udine, Via delle Scienze 206, 33100 Udine, Italy; email: david.esseni@uniud.it.}	
}

\maketitle
%
\section{Abstract}
The technological exploitation of ferroelectricity in CMOS electron devices offers new design opportunities, but also significant challenges from an integration, optimization and modelling perspective.
We here revisit the working principle and the modelling of some novel ferroelectric based devices, with an emphasis on energy efficiency and on applications to new computational paradigms.

\vspace{2mm}
\textbf{Keywords}: Negative Capacitance, Ferroelectric Tunneling Junctions, Ferroelectric FETs, Neuromorphic Computing.

\vspace{2mm}
\textbf{© 2021 IEEE. Personal use of this material is permitted. Permission from IEEE must be obtained for all other uses, in any current or future media, including reprinting/republishing this material for advertising or promotional purposes, creating new collective works, for resale or redistribution to servers or lists, or reuse of any copyrighted component of this work in other works.
}


\section{Introduction}
\label{sec:Intro}
%
The slowing down of the CMOS geometrical scaling has steered the research in electron devices for computing applications to new functionalities for innovative computational paradigms and to the energy efficiency. The energy dissipation in CMOS digital circuits can be improved by an aggressive \VDD\ scaling that, however, is hampered by the requirement of a large ratio $[$\Ion$/$\Ioff$]$. This led to the quest for transistors featuring a room temperature subthreshold swing below 60mV/dec \cite{Review_Seabaugh,Ionescu_Nature2011,Esseni_SST2017,Salahuddin_NL2008,Hongjuan_EDL2018}, and thus capable of improving $[$\Ion$/$\Ioff$]$ in near-threshold or sub-threshold circuits.
At the same time, the rise of artificial intelligence has emphasized the need for hardware platforms conceived for new computational paradigms, such as crossbar arrays for
artificial deep neural networks \cite{Ambrogio_Nature2019}, and new hybrid memristive-CMOS circuits for spike-based neuromorphic processors \cite{Yu_IEEE_Proc2018,Chicca_APL2020}.

The discovery of a robust ferroelectricity in hafnium oxides opened exciting perspectives for a exploitation of ferroelectric materials in CMOS technologies \cite{Boscke_IEDM2011,Zhou_ActaMat2015}, with prospective applications including negative capacitance transistors, non volatile memories and memristors \cite{Mikolajick_IEDM2019,Slesazeck_Nanotechnology2019}.

In this paper we review a few selected topics from this exciting research field.

%
\vspace{-1mm}
\section{Correlated domain dynamics and negative capacitance operation}
\label{sec:NCFETs}
\begin{figure}[]
	\centering
	\includegraphics[width = 1 \columnwidth]{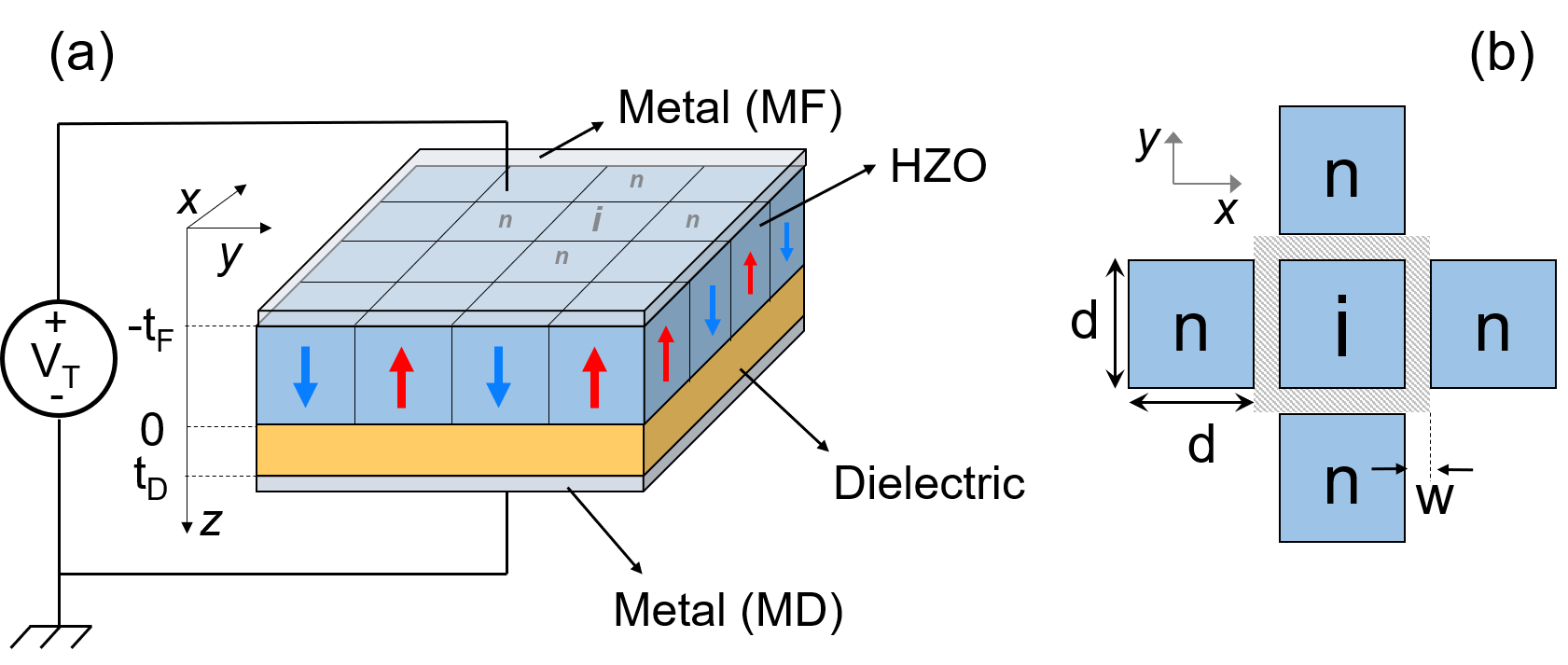}\vspace{-1mm}
	\caption{\label{Fig:MFIM_sketch} \protect \footnotesize (a) Sketch of a MFIM capacitor used as template device for simulations of both NC operation in Sec.\ref{sec:NCFETs}, and FTJs in Sec.\ref{sec:FTJs}. $MF$, $MD$ are the electrodes contacting respectively the ferroelectric and the dielectric, and $t_F$, $t_D$ are the ferroelectric and dielectric thickness.
%
%
(b) Sketch of the ferroelectric domains, where $d$ and $w$ are respectively the side of the square domains and the width of the domain-wall region entering the LGD equations \cite{Rollo_Nanoscale2020}. In this work we used $d$$=$5 nm and $w/d$$=$$0.1$.
	%
	}
\vspace{-6mm}
\end{figure}

The negative capacitance (NC) operation of ferroelectric materials and devices has been originally proposed based on a homogeneous Landau theory \cite{Salahuddin_NL2008}, and for a Metal-Ferroelectric-Insulator-Metal (MFIM) capacitor sketched in Fig.\ref{Fig:MFIM_sketch}(a). In such a single domain picture an NC stabilization condition can be derived as $(C_D$$+$$C_{F})$$<$$1/(2 |\alpha| \, t_{F})$ \cite{Hoffmann_Nanoscale2018,Rollo_Nanoscale2020}, where $\alpha$ is the anisotropy constant of the polarization ($P$) term in the ferroelectric dynamic equations, and $C_D$$=$$\varepsilon_0\varepsilon_{D}/t_D$, $C_F$$=$$\varepsilon_0\varepsilon_{F}/t_F$, with $t_D$, $\varepsilon_{D}$, $t_F$, $\varepsilon_{F}$  being the thickness and relative permittivity of the dielectric and ferroelectric materials, respectively.

%
A more thorough investigation based on the multi-domain Landau, Ginzburg, Devonshire (LGD) theory has recently pointed out that the coupling constant $k$ of the domain wall energy has an important influence on the NC stabilization \cite{Hoffmann_Nanoscale2018,Park_AdvMater2019} (see Fig.\ref{Fig:MFIM_sketch}), and a $k$ dependent stability condition was derived in \cite{Rollo_Nanoscale2020,Rollo_IEDM2019}.

From an experimental standpoint, NC observations have been first reported for ferroelectric films connected in series to an external resistance \cite{Khan_NatMat2015,Song_ScientReports_2016}, then discussed for steep slope FETs \cite{Li_IEDM2015,Lee_IEDM2016,Sharma_VLSI2017}, and more recently assessed in Metal-Ferroelectric-Insulator-Metal capacitors \cite{Hoffmann_IEDM2018,Hoffmann_Nature2019}.

\begin{figure}[]
	\centering
	\includegraphics[width = 0.7 \columnwidth]{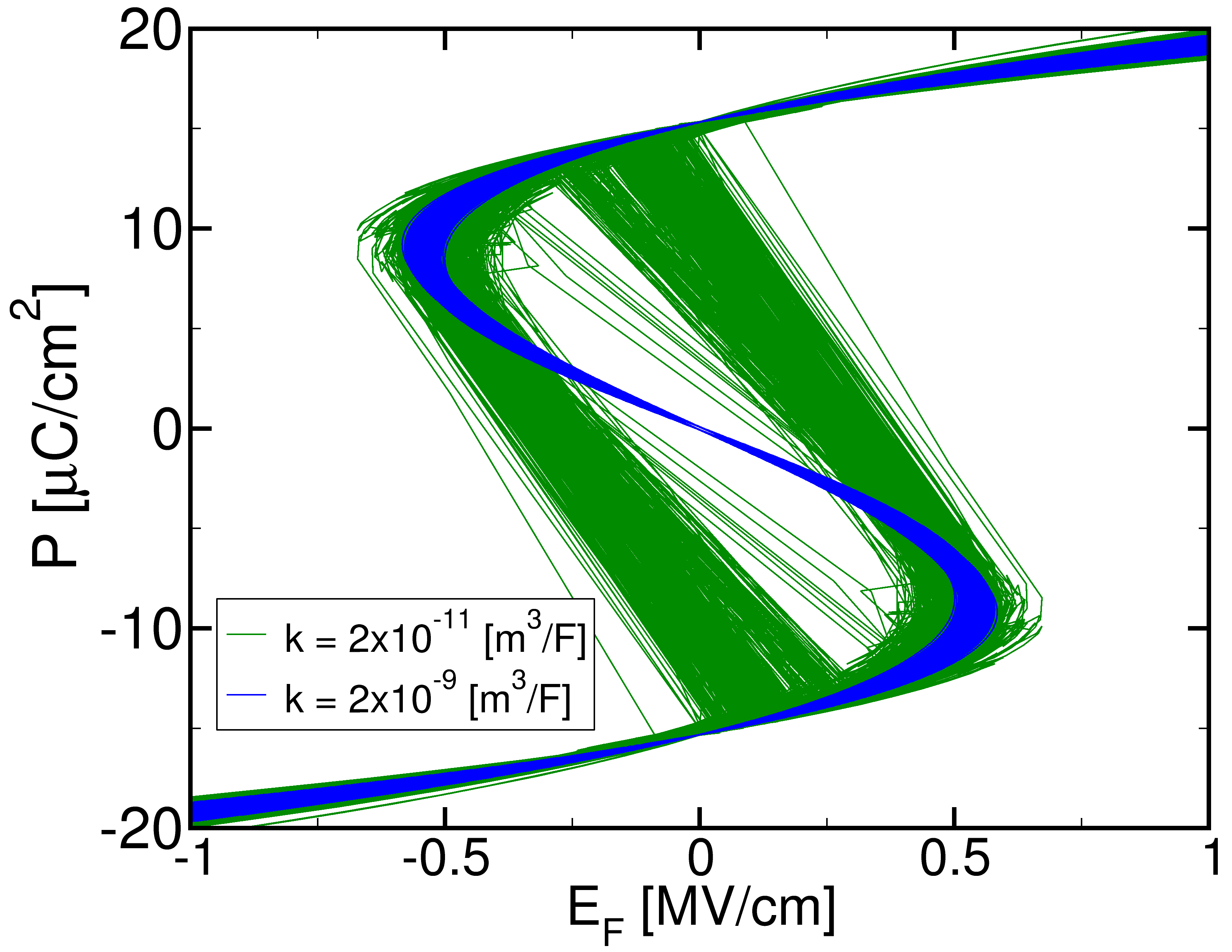}\vspace{-1mm}
	\caption{\label{Fig:NC_k_comparison} \protect \footnotesize Simulated spontaneous polarization versus ferroelectric field of each individual domain in an Hf$_{0.5}$Zr$_{0.5}$O$_2$-Ta$_2$O$_5$ MFIM structure. Simulation parameters are $\varepsilon_F=33$, $\varepsilon_D=23.5$, $t_F=11.6$ nm, $t_D=13.5$ nm, $\alpha=-4.6$$\cdot$$10^8$ m/F, $\beta=9.8$$\cdot$$10^9$ m$^5$/C$^2$/F and $\gamma=0$,
that provide good agreement with experiments
of \cite{Hoffmann_Nature2019}. The simulated device has $n_D$$=20\times20$ domains.
	}
\vspace{-5mm}
\end{figure}

From the standpoint of the multi-domain LGD equations, the NC behaviour corresponds to a strongly correlated domain dynamics as illustrated by the simulations reported in Fig.\ref{Fig:NC_k_comparison} and obtained with the modelling approach discussed in \cite{Rollo_IEDM2019,Rollo_Nanoscale2020}. For a relatively large domain wall coupling constant $k$$=$$2\cdot 10^{-9}$ m$^3$/F, the trajectories for all domains are tightly correlated, whereas for a smaller $k$ we see that the switching occurs via a domain nucleation process.

\begin{figure}[]
	\vspace{-3mm}
	\centering
	\includegraphics[width = 0.8 \columnwidth]{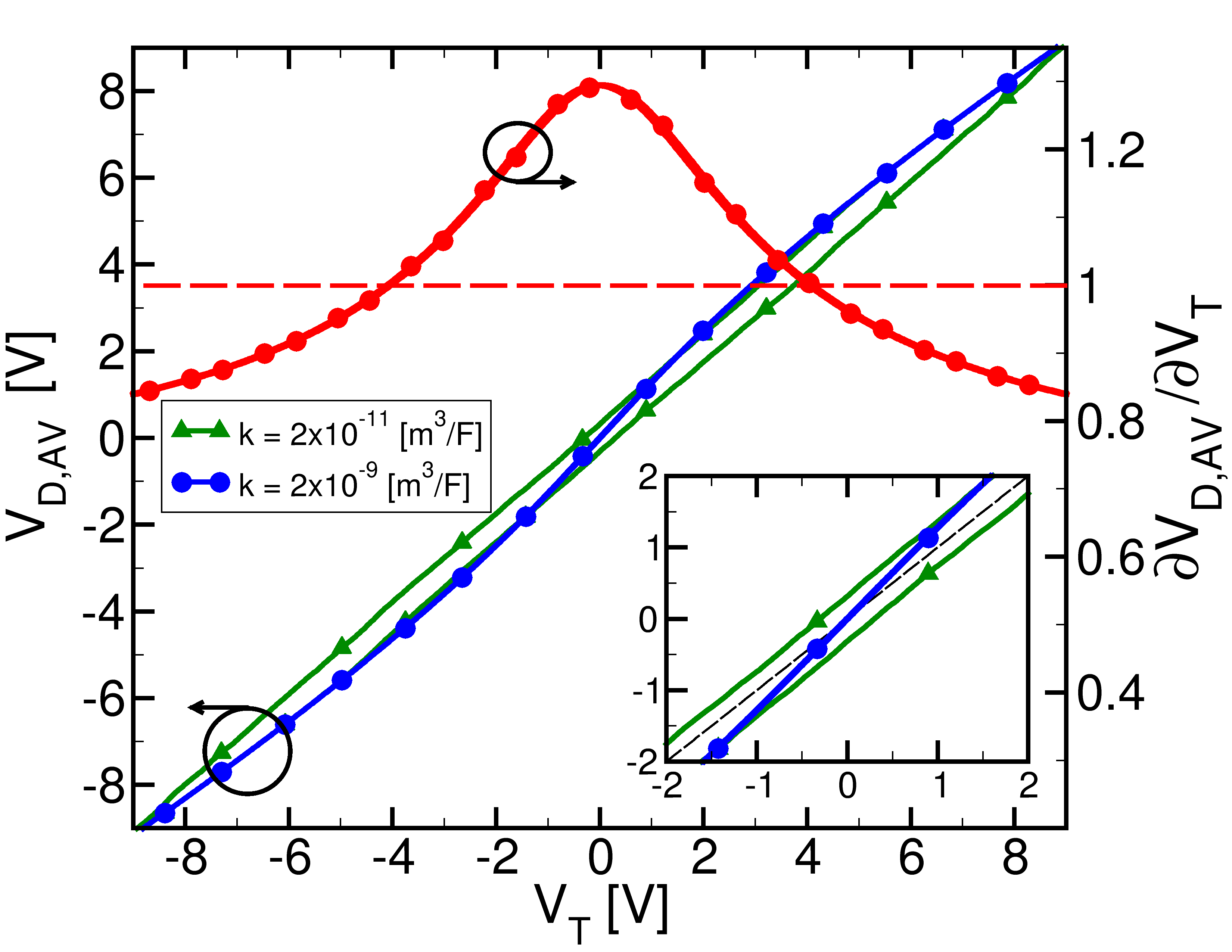}\vspace{-1mm}
	\caption{\label{Fig:VDav_k_comparison} \protect \footnotesize Voltage enhancement across the thin insulator in NC condition (blue line), compared to the hysteretic one (green line), versus applied external voltage  (same simulations reported also in Fig.\ref{Fig:NC_k_comparison}). The NC condition allows a voltage gain (red line) greater than 1 across the central $V_T$ region, between $-4$ V and $4$ V.
	}
\end{figure}

Figure \ref{Fig:VDav_k_comparison} reports the domain averaged voltage drop, $V_{D,AV}$, for the  $k$ values used in Fig.\ref{Fig:NC_k_comparison}. The correlated switching process results in a hysteresis free NC behavior and, during the $V_T$ ramp, $V_{D,AV}$ increase faster than $V_T$ itself, thus leading to a larger than one voltage gain $[$$\partial V_{D,AV}$$/$$\partial V_T$$]$, which is the original idea behind the NC exploitation in nanoscale FETs \cite{Salahuddin_NL2008}.

The possible advantages of a ferroelectric NC operation have been investigated up to a full-chip level and in a high quality  industrial technology \cite{Zoran_IEDM2017}. 
While the benefits of the NC operation on the sub-threshold characteristics of actual transistors seem only modest, design studies for NC transistors are still being actively reported \cite{Rollo_EDL2018_gm,Rollo_IEDM2018,Pentapati_TED2020}.

%
\begin{figure}[]
	\centering
	\includegraphics[width = 1.0 \columnwidth]{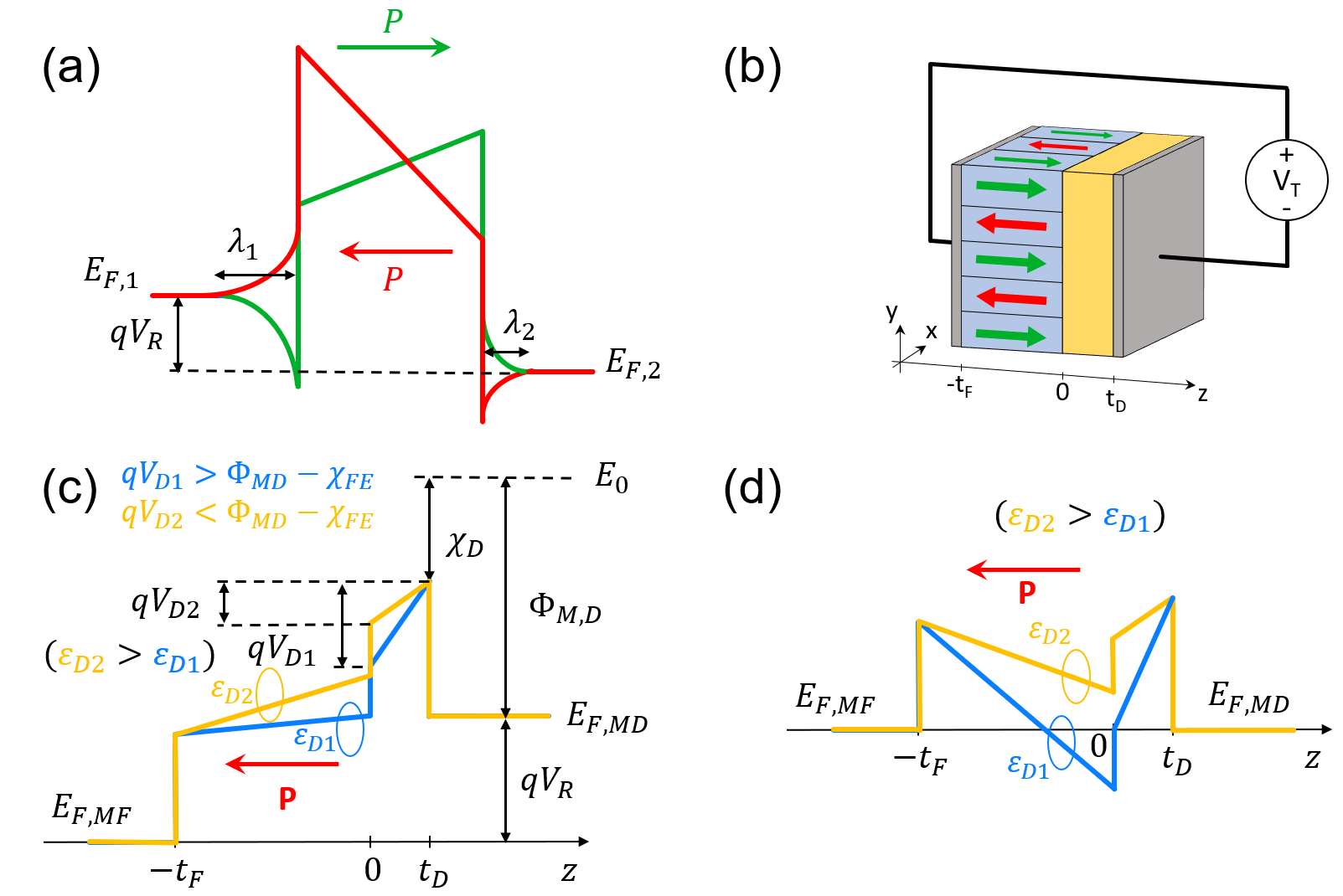}\vspace{-1mm}
	\caption{\label{Fig:Band_Diagrams} \protect \footnotesize (a) Band diagram for an MFM based FTJ, where $\lambda_1$ and $\lambda_2$ (with $\lambda_1$$>$$\lambda_2$) are the screening lengths in the metal electrodes. (b) Cross-section of a MFIM based FTJ (see also Fig.\ref{Fig:MFIM_sketch}(a)) used for simulations in Sec.\ref{sec:FTJs}. (c) Band diagram across a MFIM based FTJ during reading $V_T$$=$$V_R$. $qV_D$ should be larger than the ferroelectric tunnelling barrier $[\Phi_{M,D}$$-$$\chi_{F}]$, so that the ferroelectric conduction band profile can drop below $E_{f,MD}$. The band diagram is shown for two different dielectric constants $\varepsilon_{D2}$$>$$\varepsilon_{D1}$ of the tunnelling oxide. (d) Same as in (c) but for the retention condition at $V_T$$=$0. The depolarization field $E_{DEP}$$\approx$$P_r\left [ \varepsilon_F \left ( C_D/C_F +1 \right )\right ]^{-1}$, where $P_r$ is the remanent polarization, should be minimized, and it larger the smaller is $\varepsilon_{D}$.
%
$E_0$ and $\Phi_{M,D}$ are respectively the vacuum level and the work function of the MD electrodes. $\chi_{F}$, $\chi_{D}$ are the electron affinity of the ferroelectric and dielectric, $E_{f,MD}$, $E_{f,MF}$ are the Fermi levels of the MD and MF electrode (see also Fig.\ref{Fig:MFIM_sketch}(a)).
	}
\vspace{-7mm}
\end{figure}
\begin{figure}[b]
	\centering
	\includegraphics[width = 0.85\columnwidth]{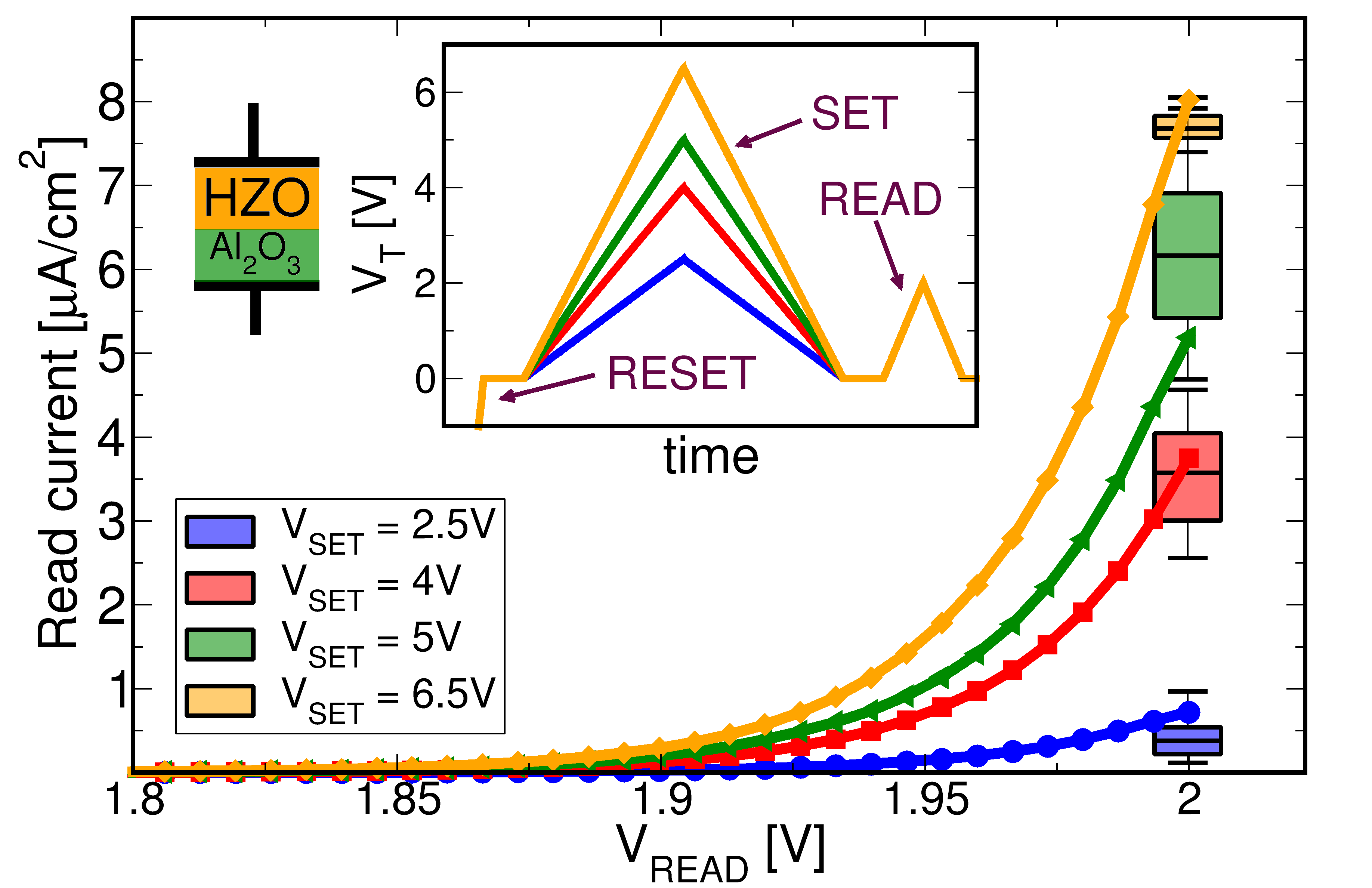}\vspace{-1mm}
	\caption{\label{Fig:Itunn_Exp_Sim} \protect \footnotesize Simulations for the read current of an HZO/Al$_2$O$_3$ FTJ ($t_F = 12$ nm, $t_D = 2$ nm) versus read voltage $V_{R}$. The box plots for experiments were inferred from the cycle to cycle variations reported Fig.3(d) of \cite{Max_JEDS2019} (device area $\approx 3.14 \cdot 10^{-4}$ cm$^{2}$). The inset shows the signal waveforms for the write and read operation. The reset voltage is $-8$ V.
}
\end{figure}

\section{Polarization dependent tunelling in FTJs}
\label{sec:FTJs}

Ferroelectric Tunnelling Junctions (FTJs) are promising candidates as highly energy efficient memristors, in fact the polarization switching is an ultra low energy mechanism to enable the potentiation and depression and, moreover, the readout impedance is very large. The originally proposed FTJ relies on a metal-ferroelectric-metal (MFM)  junction \cite{Esaki_IBM1971}, whose band diagram is sketched in Fig.\ref{Fig:Band_Diagrams}(a). The metal electrodes must have different screening lengths, $\lambda_1$ and $\lambda_2$, to obtain a polarization dependent tunnelling.

The MFIM architecture sketched in Fig.\ref{Fig:Band_Diagrams}(b), instead, separates the switching element and the tunnelling oxide, and an experimental demonstration of a Back-End-Of-Line (BEOL), four level FTJ memristor has been recently reported in \cite{Max_JEDS2019}.

A robust operation of an MFIM based FTJ requires that during reading the oxide voltage drop $V_D$ is large enough to have $qV_D$$>$$[\Phi_{M,D}$$-$$\chi_{F}]$ (see Fig.\ref{Fig:Band_Diagrams}(c)), so that tunnelling is limited by the thin dielectric layer. This can be achieved for a small enough $C_D$$=$$\varepsilon_D$$/$$t_D$ which, however, enhances the depolarization field $E_{DEP}$ destabilizing the polarization state during retention (see Fig.\ref{Fig:Band_Diagrams}(d)). It is evident that device design entails quite delicate tradeoffs that demands a sound modelling support. In this respect, Fig.\ref{Fig:Itunn_Exp_Sim} shows the simulated read current characteristics for an HZO/Al$_2$O$_3$ FTJ. A four level operation is observed and the simulated read currents are in fairly good agreement with experiments in \cite{Max_JEDS2019}.

\vspace{-2mm}
\section{Polarization dependent \IDS\ in FeFETs}
\vspace{-1mm}
\label{sec:FeFETS}
Ferroelectric FETs (FeFETs) represent an alternative device concept for neuromorphic and in-memory computing systems.
%
Several BEOL-compatible HfO based ferroelectric capacitors 
\cite{Lin_IEDM2019,Francois_IEDM2019}, 
as well as ferroelectric memristors have already been demonstrated \cite{Ni_IEDM2018, Mo_JEDSI2020,  Halter_ACSAMI2020}.

\begin{figure}[]
	\centering
	\includegraphics[width = 0.60\columnwidth]{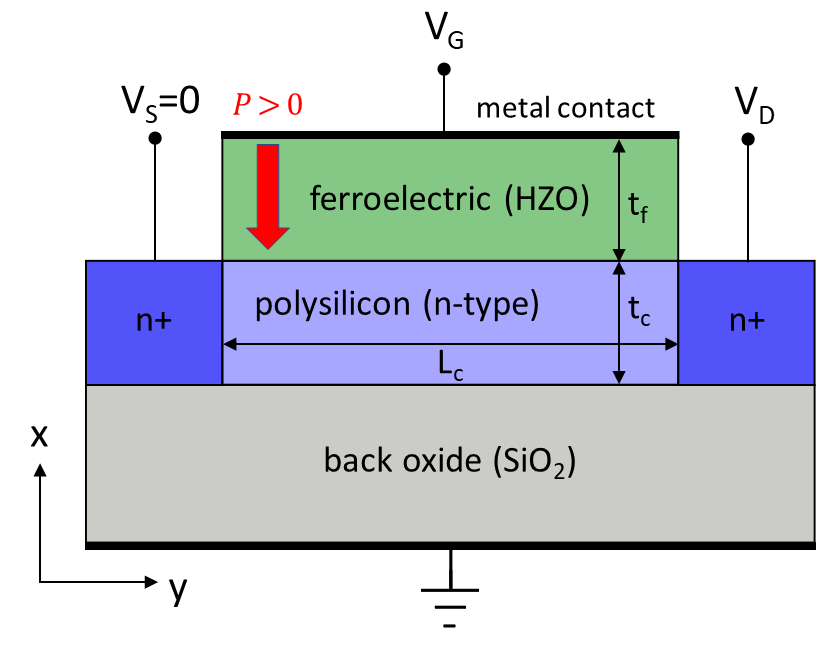}\vspace{-1mm}
	\caption{\label{Fig:FeFETDevice_structure} \protect \footnotesize
		Cross-sectional view of the simulated FeFET. The polysilicon channel is {\it n}-type with donor doping varying between N$_D=5\cdot 10^{18}$ cm$^{-3}$ and N$_D=1\cdot 10^{20}$ cm$^{-3}$ with a thickness t$_c=20$ nm and length $L_c=500$ $nm$. Source and drain regions are n-type doped (N$_D=5\cdot 10^{20}$ cm$^{-3}$). Ferroelectric  thickness is t$_f=20$ nm and Landau coefficients are reported in the caption of Fig.\ref{Fig:NC_k_comparison}. The polarization is positive when it points to the polysilicon channel.
	}
\end{figure}

\begin{figure}[]
	\centering
	\includegraphics[width = 0.8\columnwidth]{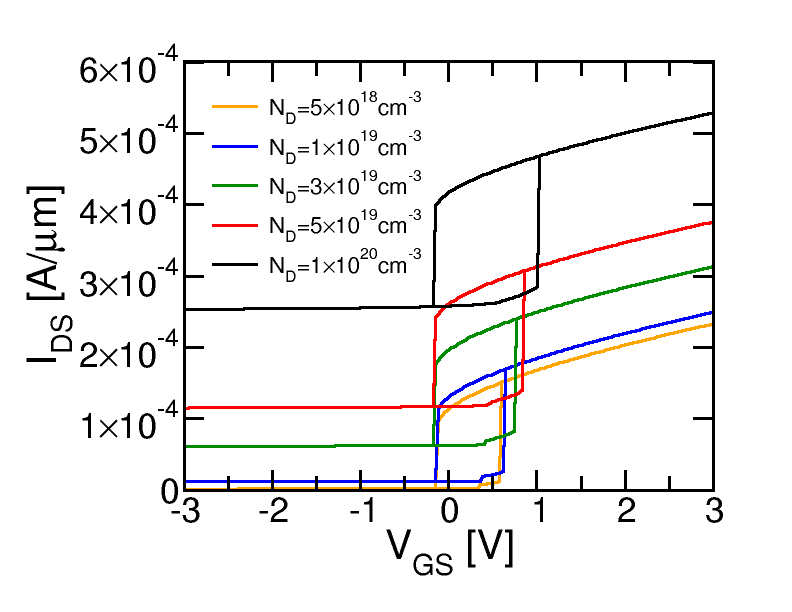}\vspace{-1mm}
	\caption{\label{Fig:FeFET_IdsVgs} \protect \footnotesize
		Quasi static I$_{DS}$-V$_{GS}$ transcharacteristic for different channel dopings for a $V_{DS}=0.05$ V. Results obtained by using a constant mobility of 100 cm$^2$/Vs.
	}
\vspace{-3mm}
\end{figure}

Figure \ref{Fig:FeFETDevice_structure} shows a 2D-FeFET with an $n$-type polysilicon active region and, as such, compatible with a BEOL integration. Simulation results are obtained using Sentaurus TCAD and are reported in Fig.\ref{Fig:FeFET_IdsVgs} for very different values of channel doping concentrations.
Because the ferroelectric layer is placed between a metal electrode and a semiconductor channel material, the semiconductor depletion can set serious limitations to the ferroelectric switching to a negative polarization (i.e. a polarization pointing to the metal contact, see Fig.\ref{Fig:FeFETDevice_structure}).
%

In this respect, a high channel doping concentration can avoid the complete depletion of the thin polysilicon layer during the switching to negative polarization. While this is a desirable behavior from a polarization switching perspective, high doping concentrations reduce the conductance modulation induced by the ferroelectric switching, as it is shown in Fig.\ref{Fig:FeFET_IdsVgs}.
%
%
On the other hand doping concentrations lower than about 10$^{19}$ cm$^{-3}$ can enlarge the difference in the read current for positive and negative polarization (see Fig.\ref{Fig:FeFET_IdsVgs}), however during the switching to negative polarization the polysilicon gets fully depleted and a minority carrier inversion is needed.
%
%
In this case the physical mechanisms for the supply of minority carriers become crucial. 

For an SRH generation of minority carriers we need small carrier lifetimes, in which case the polysilicon channel can help because the defects at the grain boundaries largely enhance the generation rates compared to mono-crystal silicon \cite{Hurley_APL1989}.
%
An alternative design option may be give by mid-gap Schottky contacts able to supply both electrons and holes, but at the cost of a larger contact resistance.
%
%
\section{Outlook and conclusions}
\vspace{-1mm}
\label{sec:Conclusions}
The ferroelectricity has introduced substantial innovation in the design of electron devices. The support of modelling and simulation is important to unleash the potentials of these materials, particularly in new applications such as hybrid
memristive-CMOS neuromorphic processing systems.
%
%
%


\noindent
\vspace{1mm}
{\bf Acknowledgements:}
This work was supported by the European Union through the BeFerroSynaptic project (GA:871737).
%
%
\vspace{-4mm}


\end{document}

%% file: macros.tex
\newcommand{\VDD}{\ensuremath{\textrm{V}_{DD}}}

\newcommand{\IDS}{\ensuremath{\textrm{I}_{DS}}}
\newcommand{\Ion}{\ensuremath{\textrm{I}_{on}}}
\newcommand{\Ioff}{\ensuremath{\textrm{I}_{off}}}

